\documentstyle[12pt]{article}
\newcommand{\bq}{\begin{equation}}
\newcommand{\ee}{\end{equation}}

\newcommand{\bi}[1]{\bibitem{#1}}
\newcommand{\fr}[2]{\frac{#1}{#2}}
\newcommand{\eps}{\varepsilon}
\newcommand{\I}{Instanton }
\newcommand{\IC}{Instanton}
\newcommand{\Is}{Instantons }
\newcommand{\YM}{Yang-Mills theory }
\newcommand{\YMC}{Yang-Mills theory}
\newcommand{\ups}{\upsilon}

\begin{document}
\pagestyle{empty}
\vspace{1.0cm}
\begin{center}{\Large \bf  Constrained Instanton and Baryon Number
Non--Conservation at High Energies }\\
\vspace{1.0cm}

{\bf  P.G.Silvestrov}\footnote{e-mail address: PSILVESTROV@INP.NSK.SU} \\
Budker Institute of Nuclear Physics, 630090 Novosibirsk, Russia

\vspace{0.5cm}

December 3, 1992

\vspace{1.5cm}

\end{center}

\begin{flushright}
{\bf BUDKERINP 92--92}
\end{flushright}

\vspace{1.0cm}

\begin{abstract}

  The total cross - section for baryon number violating processes at high
energies is usually parametrized as $\sigma_{total}
\propto\exp(\fr{4\pi}{\alpha} F(\eps))$, where $\eps =\sqrt{s}/E_0 , \,\,
E_0 = \sqrt{6} \pi m_w/\alpha$. In the present paper the third nontrivial term
of the expansion
\[
F(\eps)= -1+\fr{9}{8}\eps^{4/3} -\fr{9}{16}\eps^2 -\fr{9}{32}
\left( \fr{m_h}{m_w}\right)^2 \eps^{8/3}\log\left( \fr{1}{3\eps}\left(
\fr{2m_w}{\gamma m_h}\right)^2 \right) +O(\eps^{8/3})
\]
is obtained. The unknown corrections to $F(\eps)$ are expected to be of the
order of $\eps^{8/3}$, but have neither $(m_h/m_w)^2$, nor $\log(\eps)$
enhancement. The total cross - section is extremely sensitive to the value of
single \I action. The correction to \I action $\triangle S\sim (m\rho)^4
\log(m\rho)/g^2$ is found ($\rho$ is the \I radius). For sufficiently heavy
Higgs boson the $\rho$-dependent part of the \I action is changed drastically.
In this case even the leading contribution to $F(\eps)$, responsible for a
growth of cross - section due to the multiple production of classical
W-bosons, is changed:
\[
F(\eps)=-1+\fr{9}{8}\left( \fr{2}{3} \right)^{2/3} \eps^{4/3} +\ldots \,\, ,
\,\, \eps\ll 1\ll \eps \left( \fr{m_h}{m_w} \right)^{3/2}  \,\, .
\]

\end{abstract}

\newpage

\pagestyle{plain}
\pagenumbering{arabic}
\section{Introduction}\label{sec:1}

  A few years ago it was recognized, that the total cross-section for baryon
number violating processes, which is small like $\exp( -
4\pi/\alpha)\approx10^{-169}$ \cite{tH} (where $\alpha = g^{2}/(4\pi)$) at
low energies, may become unsuppressed at TeV energies \cite{Ring,Esp,MVV}.
It was shown \cite{KRT,AM} that the cross-section for the processes
accompanied by the multiple emission of W and H bosons has the generic form
\bq\label{eq:1}
\sigma_{total}\propto\exp\left(\fr{4\pi}{\alpha}
F(\varepsilon)\right) \ ,
\ee
where $\varepsilon = \sqrt{s}/E_0$ and $E_0 = \sqrt{6}\pi M_w/\alpha$. Up to
now the values of three nontrivial terms of $F(\varepsilon)$ expansion at
small $\varepsilon$ were claimed in the literature:
\bq\label{eq:2}
F(\varepsilon)=-1+\fr{9}{8}\eps^{4/3}-\fr{9}{16}\eps^2+
\fr{3}{16}\eps^{8/3}\log\left(\fr{c}{\eps}\right)+\ldots \ .
\ee
Here the zeroth term is the old result by 't Hooft \cite{tH}, the first term
$\sim\eps^{4/3}$ was found in refs. \cite{KRT,Zakh,Por} and the second
term $\sim\eps^2$ was obtained in refs. \cite{Khos,DiPe,Muel,ArMa}.
The last term $\sim\eps^{8/3}\log(\eps)$ was found in ref. \cite{Diak}. Authors
of \cite{Diak} have calculated the 2-loop correction to multiple W-boson
emission. They have not found the precise argument of $\log$ and guessed
$\log c=1$ for numerical estimates.

  In the present paper some new contributions to $F(\eps)$ of the order of
$\eps^{8/3}\log (\eps)$ are found which were not taken into account in
\cite{Diak}. The final result reads
\bq\label{eq:3}
F(\eps)=-1+\fr{9}{8}\eps^{4/3}-\fr{9}{16}\eps^2-
\fr{9}{32}\left(\fr{m_h}{m_w}\right)^2\eps^{8/3}\log\left(\fr{1}{3\eps}
\left(\fr{2m_w}{\gamma m_h}\right)^2\right)+O (\eps^{8/3}) \ .
\ee
Here $m_w, m_h$ are W and Higgs boson masses, $\gamma$ is the Eurler's constant
$(\log\gamma =0.5772\ldots)$. In view of the current situation with H-mass
the factor $(m_h/m_w)^2$ is not small and possibly is sufficiently
large. The corrections to (\ref{eq:3}) are of the order of $\eps^{8/3}$
but have no enhancement either $(m_h/m_w)^2$, or $\log(\eps)$ .

  The main physical question we are faced is: can the total cross-section be
large? In other words does $F(\eps)$ vanish at some $\eps\sim 1$ ? Let us see,
does one can answer this question with only a few terms of $F(\eps)$ expansion
(\ref{eq:3}). We know only one heuristic rule which allows to check the
accuracy of such a partial sum. The last term in the expansion should be
smaller than the previous one. According to this "rule" the result of
\cite{Diak}, eq. (\ref{eq:2}), still allows one to believe in the validity of
the expansion up to sufficiently large values $\eps\sim 1$. Unfortunately, the
last term of the corrected expansion (\ref{eq:3}) becomes equal to $\eps^2$
term for $m_h=2\div3 m_w$ at $\eps=0.3\div0.12$ . Moreover, in {\bf Section
\ref{sec:5}} we argue, that formula (\ref{eq:3}) may be used only at $3\eps
< \left(\fr{2m_w}{\gamma m_h}\right)^2$.

  The growth of total cross-section is dominated mostly by the multiple
production of W-bosons (the typical multiplicity is $N\sim \eps^{4/3}/g^2\gg
1$). In the leading classical approximation the birth of new W leads to the
factor
\bq\label{eq:4}
\fr{4\pi^2\rho^2}{g}U^{ab}\bar{\eta}^b_{\mu\nu}k_\nu\propto\rho^2
\ee
in baryon number violating amplitude ($\rho$ is the \I radius). Therefore
the contribution of small size \Is is
strongly suppressed in the multiparticle cross-section. In order to
perform the integration over $\rho$ it is necessary to utilize the $\rho$
dependence of the \I action, which appears in spontaneously broken Yang-Mills
theory. All the authors have used for calculation of the cross-section
the \I action of the form \cite{tH,Afl}
\bq\label{eq:5}
S_I=\fr{8\pi^2}{g^2}+\pi^2(\rho\ups)^2 \ ,
\ee
where $\ups$ is the Higgs vacuum expectation value. It is usually forgotten
that expression (\ref{eq:5}) is approximate. But it was pointed out in the
paper \cite{Afl}, that where are the corrections to the action
(\ref{eq:5}) of the form
\bq\label{eq:6}
\triangle S\sim (\rho\ups)^2(m\rho)^2\log(m\rho) \ ,
\ee
with $m$ standing for W or H mass. Just the
calculation of corrections like (\ref{eq:6}) to the Instantonic action in
the SU(2) \YM is the main subject of present paper.

Being only the approximate solution of the equations of motion, the \I in
spontaneously broken \YM should be a subject of some additional
constraint. The problem of the best choice of the constraints, which should
be used in order to introduce the collective variables, is a vital problem
for our consideration. In {\bf Section 2} we examine a possibility to
distinguish consistently between the collective and quantum variables for
the approximate \IC. The result is rather pessimistic. Only the negative
restrictions on the possible choice of constraint may be formulated. The
direction in the functional space associated with the collective variable
should not be {\it orthogonal} to the unperturbed \I zero mode, but also
should not {\it coincide} with (or be {\it very close} to) the zero mode.
Only a few terms of the \I action expansion over $(m\rho)$ are constraint
independent. Only those model independent terms may be used undoubtedly for
the calculation of $\sigma_{total}$.

  In {\bf Section 3} we investigate the shape of the constrained \IC.

  In {\bf Section 4} the corrections to $S_I$ of the type (\ref{eq:6}) are
calculated. There appears two large parameters, which allow to classify
various contributions to $\triangle S$. These are $(m_h/m_w)^2$ and
$\log(m\rho)$. Only the contributions to $\triangle S$ of the form $\sim
(\rho\ups)^2 (m_w\rho)^2$ which have neither $(m_h/m_w)^2$, nor $log(m\rho)$
enhancement may be constraint dependent. Therefore we
cannot calculate the contributions to $F(\eps)$ (\ref{eq:3}) of the order
of $O(\eps^{8/3})$ which are not enhanced by either $(m_h/m_w)^2$, or
$\log(\eps)$.

  An additional correction to $F(\eps)$ of the interesting type
comes from the Higgs boson production (see {\bf Section 6}). The multiple
Higgs boson production in the
limit $m_h=0$ contribute to $\eps^2$ term of $F(\eps)$ expansion. But taking
into account the finite $m_h$ leads to the correction
$\sim(m_h/m_w)^2\eps^{8/3} \log(\eps)$ .

  It seems very attractive to sum up all the corrections to $F(\eps)$
enhanced by the power of $(m_h/m_w)$ exactly. In {\bf Section \ref{sec:a}}
the \I
action for the case $(m_w\rho)\ll 1\ll (m_h\rho)$ is calculated. The value
of $S_I$ in this case differs from the well known result (\ref{eq:5}) of
refs. \cite{tH,Afl}. As a result even the term in $F(\eps)$ which appears
due to the multiple massless W-boson production differs in this case from that
found in refs. \cite{KRT,Zakh,Por}
\begin{eqnarray}\label{eq:c1}
&F(\eps)={\displaystyle -1+\fr{9}{8}\left( \fr{2}{3}\right)^{2/3} \eps^{4/3}
+\ldots} \,\, , \\
&{\displaystyle \eps\ll 1 \,\, , \,\,
\eps\left( \fr{m_h}{m_w}\right)^{3/2} \gg 1}  \,\, . \nonumber
\end{eqnarray}

  In last two years some attempts were made to use the \I for calculation of
the multiparticle cross-section in massive $\phi^4$ theory \cite{ShM}. The
authors of \cite{ShM} have used $S_I$ -- the \I action which was found in
\cite{Afl}. We tried\footnote{this part of the work was done together with
M.E. Pospelov.} to improve (see {\bf Section \ref{sec:Phi}}) the
calculation of \I action in massive $\phi^4$
theory and to our surprise got the result different from that of \cite{Afl}.

\section{The choice of the constraint}\label{sec:2}

  Consider SU(2) \YM coupled to a doublet of Higgs complex fields
\begin{eqnarray}\label{eq:7}
S&=&\int \left\{\ \fr{1}{4}F^a_{\mu\nu}F^a_{\mu\nu} +
\fr{1}{2}\overline{(D_\mu\Phi)}(D_\mu\Phi) +
\fr{\lambda}{8}(\overline{\Phi}\Phi-\ups^2)^2 \right\}\ d^4r \ , \nonumber\\
D_\mu&=&\partial_\mu-\fr{ig}{2}\sigma^aA^a_\mu \ .
\end{eqnarray}
If one forget about the last ($\sim \lambda$) term in the action, there may
be found an \I
\bq\label{eq:8}
A^a_{\mu 0} = \fr{2}{g} U^{ab}\bar{\eta}^b_{\mu\nu}\fr{r_\nu\rho^2}
{r^2(r^2+\rho^2)} \quad , \quad \Phi_0 =\fr{r}{\sqrt{r^2+\rho^2}}
\left( \begin{array}{c}
   0 \\ \ups
\end{array} \right) \ .
\ee
But for the full theory (\ref{eq:7}) no nontrivial extrema of the action
exist, as can be seen from the transformation
\bq\label{eq:9}
A^a_\mu(r) \longrightarrow aA^a_\mu(ar) \quad , \quad \Phi(r)
\longrightarrow \Phi(ar) \ .
\ee
Only the zero size \I can be the exact extremum of (\ref{eq:7}).
Nevertheless, the functional integral for the amplitudes of processes
changing the baryon number
should be saturated by the \I -- like configurations very close to \I
(\ref{eq:8}) of a very small size $m\rho\ll 1$. These configurations are only
the approximate extrema of (\ref{eq:7}) since their action $S_I$
depends slightly on $\rho$ (the constrained \Is \cite{Afl,FY}).

  Although at $m\rho\ll 1$ the action of the approximate \I depends on
$\rho$ only slightly, the field configuration itself changes drastically with
the change of Instanton radius. It means that while $\rho$ is not associated
with the exact zero mode, the integration over the size of Instanton should be
performed with the collective variables method. One introduces the
constraint into the functional integral, which allows to integrate over all
the short-wave quantum variables before the integration over $\rho$
\begin{eqnarray}\label{eq:cons}
& &\delta(\langle A\mid W\rangle -\tau )\, d(\langle A\mid W\rangle
-\tau ) \,\, ,  \\
& &\langle A\mid W\rangle \equiv \int A_\mu^a W_\mu^a(\vec{r})d^4r \,\, .
\nonumber
\end{eqnarray}
Here $\tau =\tau (\rho)$. The vector $W^a_\mu(r)$ may also,
generally speaking, depend on $\tau$. After we have imposed the constraint
(\ref{eq:cons}), the functional integration is to be done
over the hyper--planes orthogonal to the constraint function $W^a_\mu(r)$. We
would
naturally call the constrained \I the configuration which minimizes the
action for a given value of collective variable $\tau$. The equation of
motion for the constrained \I reads
\bq\label{eq:em}
\fr{\delta S}{\delta A^a_\mu} = \zeta W^a_\mu \,\, ,
\ee
where $\zeta$ is a Lagrange multiplier, which allows to satisfy the
constraint (\ref{eq:cons}).

  Of course, one can try not to find any extremum of the action, but work
with the "wrong" configuration (\ref{eq:8}). The linear terms, which appear
after expansion of the action near the wrong minimum, may be treated
perturbatively. But doing so, one has to be convinced that the "wrong"
minimum (\ref{eq:8}) actually differs very slightly from the exact solution
of the constrained equation of motion (\ref{eq:em}). This is not evident even
if we are dealing with simple few particle amplitudes. In this case the usual
correction to \I action $\triangle S\sim (m_w\rho)^2/g^2\sim 1$ is comparable
with the one loop quantum corrections. But correction to the action which
comes from a naive substitution of the unperturbed \I (\ref{eq:8}) into the
last ($\sim\lambda$) term of the action (\ref{eq:7}) diverges logarithmically,
thus reflecting the fact that the unperturbed Instantonic solution should be
modified. In this paper we will be interested (see Sec. \ref{sec:5}) in the
processes of a huge multiplicity $N\sim\eps^{4/3}/g^2$. In this case the
corrections to \I action are large like $\sim g^{-2}$ even though the \I is
small $(m\rho)^2\sim\eps^{4/3}\ll 1$.

  Since the constrained \I evidently depends (see (\ref{eq:em})) on the vector
$W^a_\mu(r)$, the "best" choice of $W^a_\mu$ seems to be the vital problem
for our consideration. On the general grounds one can only state, that the
result of the exact calculation of the functional integral (if it is possible
to perform) should be constraint independent. From the "naive" analysis of the
role of zero modes in Instantonic phenomena it seems almost evident that, as
our constrained \I looks very similar to the unperturbed \I, the vector
$W^a_\mu(r)$ for the "best" constraint should be very close to the \I zero
mode. The main aim of this Section is to show that such a "naive" solution of
the best constraint problem is {\em wrong}.

  We introduce the constraint (\ref{eq:cons}) in order to prohibit the
integration along some dangerous directions in the functional space. While
doing so we suppose, that all the rest integrals are well convergent and may
be treated as quantum corrections in one--loop approximation. Therefore the
only {\it a priori} requirement to our constraint is that it should effectively
suppress the integration along all "dangerous" directions in the functional
space, thus providing us with reasonable quantum corrections.

  The problem of an appropriate choice of the constraint first appears for
the functional integral calculation around the exact \I of pure \YMC, which we
overview before considering the approximate \IC. The action of SU(2) theory
does not change under dilatation, translation and isotopic spin rotation of
\IC. Therefore in order to calculate the functional integral (though to one
loop) one should prohibit in some way the integration along a few directions
in the functional space. To this end we introduce eight $\delta$-functions in
the integral. After that the Euclidean functional integral takes the form
\bq\label{eq:10}
Z= \int \prod_{i=1}^{8} \delta \biggl(\langle W_i \mid A-A_I\rangle
\biggr)d\langle W_i\mid A-A_I\rangle
\exp \Biggl\{ -S_{YM}-S_{fix}-S_{ghost} \Biggr\} DAD\phi_{ghost} \ ,
\ee
where $S_{fix}$ may be used e.g. \cite{tH}
\[
S_{fix}=\int \fr{1}{2}(D^{cl}_\mu (A^a_\mu - A^a_{\mu I}))^2 d^4r \ .\]
We suppose that $D_\mu^{cl} W_\mu =0$, otherwise the preexponential factor
in (\ref{eq:10}) should be modified slightly. It is easy to transform the
integration over $d\langle W_i \mid A-A_I\rangle$ to that over the usual
collective variables $\rho$, $\vec{r_0}$ and angles of the isotopic spin
rotation matrix $U_{ab}$. At least until we consider the pure \YM there is a
great freedom in the choice of the "constraints" $W^a_{\mu i}$. The explicit
form of the Instantonic solution (\ref{eq:8}) does not depend on the
constraint.
The quantum correction to the two point Green function, for example, depends
on $ W^a_{\mu i} $. But in the final result for any observable value the
constraint dependence should disappear to any order of the coupling constant.

  Nevertheless some important restrictions may be formulated which the allowed
constraints should satisfy, if one wants to have reasonable quantum
corrections.

  {\large\bf I.} The functions $W^a_{\mu i}$ should not be orthogonal to the
\I zero modes $W^{a(0)}_{\mu i}$. More precisely the matrix $J_{ij}=\langle
W_i\mid W^{(0)}_j\rangle$ should not be degenerate $\Bigl( detJ \neq 0\Bigr)
$.

  {\large\bf II.} The functions $W^a_{\mu i}$ should not coincide with (and
should not be very close to) the exact zero modes $W^{a(0)}_{\mu i}$. This
requirement, not so obvious as {\large\bf I.}, needs additional comments.

  Let us consider for example the Green function $G^{ab}_{\mu\nu}$. Suppose,
we do use the exact zero modes $W^{a(0)}_{\mu\nu}$ as the constraint functions.
Then Green function should satisfy the equation \cite{Br}:
\begin {eqnarray}\label{eq:14}
M^{ab}_{\mu\sigma} G^{bc}_{\sigma\nu}&=&\delta^{ac} \delta_{\mu\nu} \delta
(x-y)
-\sum_{i=1}^{8}W^{a(0)}_{\mu i}(x)W^{c(0)}_{\nu i}(y) \nonumber\\ \ \\
M^{ab}_{\mu\nu}&=&-\delta_{\mu\nu}D^{ac}_\sigma D^{cb}_\sigma -
2g\eps^{acb}F^c_{\mu\nu} \ . \nonumber
\end{eqnarray}
The normalized zero modes $W^{a(0)}_{\mu i}$ are themselves the
eigenfunctions of $M^{ab}_{\mu\nu}$ having zero eigenvalues. Let
$\Psi^a_{\mu k}$ be the set of eigenfunctions of the operator
$M^{ab}_{\mu\nu}$ having eigenvalue $\eps_k$. The formal solution to the
equation (\ref{eq:14}) is
\bq\label{eq:15}
G^{ab}_{\mu\nu} = \sum_{k=9}^{\infty} \fr{\Psi^a_{\mu k}(x) \Psi^b_{\nu
k}(y)} {\eps_k} \ .
\ee
Formula (\ref{eq:10}) with our choice of constraint $W_i \equiv W^{(0)}_i$
implies that eight zero modes $W^{a(0)}_{\mu i}$ (the eight terms
$k=1,...,8$) should be omitted in the sum (\ref{eq:15}). But in massless
theories like pure Yang-Mills there exists an infinite number of operator
$M^{ab}_{\mu\nu}$ eigenfunctions having arbitrary small eigenvalues. As we will
see below, the summation over these "soft" fluctuations leads to a
divergence of the Green function (\ref{eq:15}). The "soft" eigenmodes $\Psi_k$
of $M^{ab}_{\mu\nu}$ (those of $\eps_k\ll\rho^{-2}$) can be found explicitly.
But we can prove that the Green function (\ref{eq:15}) is infinite without an
explicit summation. At large $x,y$ the eq. (\ref{eq:14}) is simplified
drastically
\begin{eqnarray}\label{eq:16}
&&-\triangle G^{ab}_{\mu\nu}=\delta^{ab}\delta_{\mu\nu}\delta (x-y) \ , \\
&&G^{ab}_{\mu\nu}=\delta^{ab}\delta_{\mu\nu}\fr{1}{4\pi^2(x-y)^2}\,\,\,\, ,
\,\,\,\, x,y\gg\rho\,\, . \nonumber
\end{eqnarray}
Because we have used the exact zero modes $W^{(0)}_i$ as constraint functions
in the functional integral (\ref{eq:10}), the Green function
(\ref{eq:15}) should be orthogonal to zero modes.

  Throughout the paper we are only interested in the explicit expression for
dilatational zero mode
\bq\label{eq:13}
W^{a(0)}_{\mu D} \sim\fr{\partial A^a_{\mu I}}{\partial\rho} \sim
U^{ab}\fr{\bar{\eta}^b_{\mu\nu}r_\nu}{(r^2 +\rho^2)^2} \,\, .
\ee

  It is seen
immediately from eqs. (\ref{eq:16}),(\ref{eq:13}) that $G^{ab}_{\mu\nu}$
averaged over dilatational zero mode $W^{a(0)}_{\mu D}$ is
\bq\label{eq:17}
\int W^{a(0)}_{\mu D}(x)G^{ab}_{\mu\nu}(x,y)W^{b(0)}_{\nu D}(y) d^4xd^4y =
\infty \,\,.
\ee
Any finite solution of eq. (\ref{eq:14}) has asymptotic (\ref{eq:16})
and thus is not orthogonal to the dilatational zero mode. The solution of
(\ref{eq:14}) can be made orthogonal to zero modes by a transformation
\bq
\tilde{G}^{ab}_{\mu\nu}=G^{ab}_{\mu\nu}+\alpha W^{a(0)}_{\mu D}(x)
W^{b(0)}_{\nu D}(y)
\ee
with suitable $\alpha$. But due to (\ref{eq:17}) one has to add the zero
modes with infinitely large weight $\alpha$ to the finite solution of
(\ref{eq:14}) in order to get the orthogonalized Green function.

  We have used the $\delta$-functions in the functional integral
(\ref{eq:10}) in order to avoid the integration along the most dangerous
directions in the functional space. We call "dangerous" those directions in
the functional space along which the Gaussian integration cannot be used.
But there turns out to be an infinite number of such directions among the
eigenmodes of the operator $M$ (eq. (\ref{eq:14})). These are not only the
eight zero modes, but also the infinite number of soft, longwave excitations.
If we chose zero modes as the constraint functions in (\ref{eq:10}), then we
leave unaffected (due to orthogonality of Hermitian operator $M$ eigenvalues)
the integral over all the rest dangerous modes. In order to affect the integral
along infinite number of modes with only a few constraints one should choose
the functions $W^a_{\mu i}$ to be different from any eigenmode of the
operator $M^{ab}_{\mu\nu}$ (eq. (\ref{eq:14})).

  In the paper \cite{Br} the formal expression for the Green function
(\ref{eq:14}) was found. But if one calculate explicitly the integral in
this expression, the Green function become infinite. In ref. \cite{LY} the
finite well-defined Green function for \YM gauge fields with constraints
decreasing much faster than the exact zero modes was found.

  Now we can turn back to the \I in spontaneously broken \YM (\ref{eq:7}).
As we have said before only a small size \I ($m_w\rho\ll 1,m_h\rho\ll 1$) may
be treated semiclassically. Like in pure \YMC, one need eight
$\delta$--functions in the functional integral (\ref{eq:10}) in order to
describe the \I collective variables. At $m\rho\ll 1$ not only \I itself,
but also the quantum fluctuations around it should differ very slightly from
that in the pure Yang-Mills case. Therefore all our arguments concerning a
possible choice of constraints are still valid. The
constrained \I is a solution of eq. (\ref{eq:em}), where the vector
$W^a_\mu$ is taken from the constraint associated with dilatational zero
mode. The solution of (\ref{eq:em}) differs
slightly from the \I of pure \YM and depends,
though also slightly, on the explicit choice of the constraint. Like there
is no the best choice of constraint, there is also no the best
choice of the \I in spontaneously broken \YMC. Of course, if we really
discuss a very small \IC, all the constraint dependence should disappear in the
final results to any order of perturbation theory (see for example the proof
of constraint indepedence of the quantum corrections to total
cross-section for baryon number violating processes \cite{DiPe,Diak}).

  The problem we have discussed looks very similar to the problem of
definition of the \I--Antiinstanton (I--A) configuration. The I--A pair is
only an approximate minimum of the action. In order to describe such a
configuration one should solve the constrained equation of the type
(\ref{eq:em}). Thus
if we want to find exactly the I--A configuration, we are again faced with
the problem of the best constraint. There exist various approaches to the
problem of the "best" choice of the I--A configuration \cite{BY,meK}. But
for pseudoparticles far separated from each other both approaches quoted above
turn out to be very close. In this case the only particular choice of
constraint functions
which appears in the right hand side of the eq. (\ref{eq:em}) is the linear
combination of the single pseudoparticle zero modes. But as we learned, just
that choice of constraint leads to unreliable quantum corrections, at least
for the \YMC. Like for the \I in spontaneously broken \YMC,
there is no "best" constraint for I--A pair in pure \YMC. Just a few
terms of I--A interaction expansion over $R^{-1}$ ($R$ is the distance
between pseudoparticles) are model independent, for example the $R^{-4}$
term for \YMC. The problem of the exact determination of I--A interaction
seems to have no sense at all.

  In view of such an impossibility to define exactly the I--A potential the
approach of ref. \cite{Bal} looks very instructive. The author of
\cite{Bal} have found the parameters of singularity caused by I--A
contribution to the Borel transform of $R_{e^+ e^-} \rightarrow
hadrons$. In order to find the type and strength of singularity one needs
just a few terms of I--A interaction. Such singularities of the Borel
transform are responsible for the large orders of perturbation theory
behaviour. On the other hand, the smooth part of the Borel function is
mostly affected by the first terms of perturbative expansion. Thus the
problem of the calculation of the short range part of I--A interaction
should be replaced by the problem of exact calculation of a few first terms
in the perturbation theory expansion.

\section{Computing the Constrained Instanton}

  Thus the constrained \I in SU(2) \YM is a solution of the equation (see
eq. (\ref{eq:em}))
\bq\label{eq:20}
\left\{
\begin{array}{ll}
{\displaystyle \fr{\delta S} {\delta A^a_\mu}}
 = -D^{ac}_{\nu}G^c_{\nu\mu}+
{\displaystyle \fr{ig}{4}} \{
\overline{\Phi}\sigma^a D_\mu\Phi -\overline{(D_\mu\Phi)}\sigma^a\Phi \} =
\zeta
W^a_\mu \\ \ \\
{\displaystyle 2\fr{\delta S}{\delta\overline{\Phi}}
 = -D^2\Phi + \fr{\lambda}{2}\Phi
(\overline{\Phi}\Phi-\ups^2)=0 \,\,\,\,\, .}
\end{array}
\right.
\ee
Here $\zeta$ is the Lagrange multiplier and $W^a_\mu$ is the constraint
function. We suppose that $W^a_\mu$ at $r>\rho$ ($\rho$ is the \I size)
goes to zero faster than any other function which we consider. The well known
\cite{tH} approximate solution of (\ref{eq:20}) is given by (\ref{eq:8}). In
order to solve the system (\ref{eq:20}) let us choose the ansatz
\bq\label{eq:22}
A^a_\mu = \fr{2}{g} U^{ab}\bar{\eta}^b_{\mu\nu}r_\nu f(r) \,\,\, , \,\,\,
W^a_\mu = \fr{2}{g} U^{ab}\bar{\eta}^b_{\mu\nu}r_\nu w(r) \,\,\, , \,\,\,
\Phi = \phi(r)
\left( \begin{array}{c}
 0 \\ \ups
\end{array} \right) \,\, .
\ee
It is easily seen, that $D_\mu^{cl}A_\mu \equiv 0$.
Now instead of (\ref{eq:20}) one gets (see for example ref. \cite{tH} for
properties of $\eta$ -- symbols):
\bq\label{eq:23}
\left\{ \begin{array}{ll}
{\displaystyle -f'' - \fr{5}{r} f'
- 12f^2 +8r^2f^3 + m_w^2\phi^2f =
\zeta w } \\ \ \\
{\displaystyle -\phi'' - \fr{3}{r}\phi' +
3r^2f^2\phi + \fr{m_h^2}{2}\phi(\phi^2-1)
= 0 \,\,\,\,\, .}
\end{array}
\right.
\ee
Here $m_w =gv/2 , m_h=\sqrt{\lambda}\ups .$ In this Section we discuss only
very small \Is $m_w\rho\ll 1 , m_h\rho\ll 1$. Therefore one can explore two
different asymptotic expansions for the solution of (\ref{eq:23}). At
$m_w r\ll 1 , m_h r\ll 1$ one has
\bq\label{eq:24}
\left\{ \begin{array}{ll}
{\displaystyle f = \fr{\rho^2}{r^2(r^2+\rho^2)} +
m_w^2 f_1(r/ \rho )
+ m_w^4 \rho^2 f_2(r/ \rho ) + \ldots } \\ \, \\
{\displaystyle \phi = \fr{r}{\sqrt{r^2+\rho^2}} + \ldots \,\,\,\, .}
\end{array}
\right.
\ee
On the other hand at $r\gg\rho$ one can use the expansion
\bq\label{eq:25}
\left\{ \begin{array}{ll}
{\displaystyle f=\fr{(m_w\rho)^2}{r^2} \left[ \fr{K_2(m_w r)}{2} +
(m_w\rho)^2 F_1(m_w r) + \ldots \right] } \\ \, \\
{\displaystyle \phi = 1 - \fr{m_h\rho^2}{r} \left[ \fr{K_1(m_hr)}{2} +
\ldots \right] \,\,\,\,\,\, , }
\end{array}
\right.
\ee
where $K_1$ and $K_2$ are the McDonald functions, $F_1$ is an unknown
function. It is our luck that at $\rho\ll r\ll m^{-1}$ both expansions
(\ref{eq:24}) and (\ref{eq:25}) are valid. Due to that we have found
explicitly the coefficients at McDonald functions in (\ref{eq:25}) from a
comparison with (\ref{eq:24}) at $\rho\ll r\ll m^{-1}$. The small $x$
asymptotic of McDonald functions reads
\begin{eqnarray}\label{eq:26}
K_1(x)&=&\fr{1}{x}+\fr{x}{2}\left[ \log{\left( \fr{\gamma x}{2} \right)}
- \fr{1}{2} \right] + \ldots \,\,\,\, , \\
K_2(x)&=&\fr{2}{x^2}-\fr{1}{2} + \ldots \,\,\,\, , \nonumber
\end{eqnarray}
where $\log\gamma = 0.5772\ldots$ is the Eurler's constant.

  The system of equations (\ref{eq:23}) has an unique localized solution for
any given value of the Lagrange multiplier $\zeta$. One may vary the value
of $\zeta$ in order to get the solution satisfying the given constraint
(\ref{eq:cons}). In other words, one can choose the Lagrange multiplier
$\zeta$ in order to get the Instantonic solution of a given size $\rho$.
Still we have not given the exact definition of $\rho$ --- the size of
constrained \IC. One can call the "Instanton of a size $\rho$" the solution
of eq. (\ref{eq:23}), which coincides with the \I of pure \YM at
$r\ll\rho$. This means $f_1(0) =0, f_2(0) =0, ...$ (see (\ref{eq:24})). But
for our purpose, to estimate the total cross-section for baryon number
violating processes, it is more convenient to use another definition of
constrained \IC. We will call the \I of a size $\rho$ in the spontaneously
broken \YM the solution of eq. (\ref{eq:23}) which at large $r$ behaves like
\bq\label{eq:32}
A_\mu^a(r\gg m^{-1})\equiv \fr{2}{g} U^{ab}\bar{\eta}^b_{\mu\nu} r_\nu
\fr{(m_w \rho)^2}{2r^2} K_2(m_w r) \ .
\ee
This mean, we require, that the function $F_1(m_w r)$ in eq. (\ref{eq:25})
should go to zero at large $r$ much faster than the McDonald function
$K_2(m_w r)$.

  Substitution of (\ref{eq:24}) into (\ref{eq:23}) provides us with the
equation for $f_1$
\bq\label{eq:27}
-\fr{d^2}{dr^2}f_1 - \fr{5}{r}\fr{d}{dr}f_1 -
\fr{24\rho^2}{(r^2+\rho^2)^2}f_1 = \fr{\zeta}{m_w^2}w -
\fr{\rho^2}{(r^2+\rho^2)^2}
\ee
The solution to this equation regular at small $r$ reads
\begin{eqnarray}\label{eq:28}
f_1\left(\fr{r}{\rho}\right)=&-&\fr{1}{(r^2+\rho^2)^2}\int_{0}^{r}
\fr{(x^2+\rho^2)^4}{x^5}
dx\int_{0}^{x}\fr{y^5}{(y^2+\rho^2)^2}\left[ \fr{\zeta}{m_w^2}w-
\fr{\rho^2}{(y^2+\rho^2)^2} \right] dy + \\
&+&\alpha\fr{\rho^4}{(r^2+\rho^2)^2}
\,\,\, . \nonumber
\end{eqnarray}
Here a solution of homogeneous equation is added with an arbitrary
weight $\alpha$. The $\alpha$ is the only parameter in (\ref{eq:28}), which
may depend on $(m_w\rho)$ due to the boundary conditions at large $r$.
For arbitrary Lagrange multiplier $\zeta$ the solution of
eq. (\ref{eq:24}) at large $r$ has the form
\bq\label{eq:29}
f=A\fr{K_2(m_w r)}{r^2} + B\fr{I_2(m_w r)}{r^2} \ ,
\ee
where $I_2$ is the Veber function, which grows exponentially with $m_w
r\gg 1$, and behaves like $(m_w r)^2$ at $m_w r\ll 1$. The Lagrange
multiplier $\zeta$ is just chosen to obtain
$B\equiv0$. From (\ref{eq:25}), (\ref{eq:26}) and(\ref{eq:28}) one gets
\bq\label{eq:30}
\zeta\int_{0}^{\infty}\fr{r^5}{(r^2+\rho^2)^2}w(r)dr=\fr{m_w^2}{6}
\ .
\ee
It is seen now from (\ref{eq:28}) that only short range $(r\sim\rho)$
behaviour of function $f_1$ depends on the choice of constraint function $w$.

  Equation for $F_1(m_w r)$ is easy to found from eqs.
(\ref{eq:23}) and (\ref{eq:25})
\bq\label{eq:F1}
-\fr{d^2}{dx^2}F_1-\fr{1}{x}\fr{d}{dx}F_1 +\fr{4}{x^2}F_1 +F_1= \fr{3}{x^2}
K_2^2(x) +\fr{m_h}{m_w}\fr{1}{x} K_1(\fr{m_h}{m_w}x) K_2(x) \,\,\, .
\ee
 It is only important for us, that this equation has no small parameters
either $(m_w\rho)$, or $(m_h\rho)$. The function $F_1(x)$ may depend only on
the ratio $(m_w/m_h)$. The explicit expression for $F_1(x)$ may be found
in terms of the integrals of the Bessel functions. But in the large logarithm
approximation we do not need such the explicit formula. The first two terms
of $F_1$ expansion at small $x$ are
\bq\label{eq:33}
F_1(x)=-\fr{1}{x^4}-\fr{1}{x^2}[\log{(x)}+c_1]\ .
\ee
For inconvenient definition of the size
$\rho$ of constrained \I large logarithms $\log (m\rho)$
may appear in the constant $c_1$ due to the boundary
condition for $F_1(x)$ at small $x$. But we have fixed the value of $\rho$
by only the long range behaviour of the \I (\ref{eq:32}). The only demand
that $F_1(x)$ should decrease at large $x$ much faster than the McDonald
function $K_2(x)$ allows to fix uniquely the function $F_1$. Now the large
distance correction to \I field $F_1(x)$ cannot depend on either
$(m_w\rho)$, or $(m_h\rho)$ even due to the boundary conditions. Thus the
constant in eq. (\ref{eq:33}) should be $c_1\sim 1$. With use of
explicit formula for $F_1(x)$ one can find the value of $c_1$. But in the
large logarithm approximation it is enough to know that $c_1\sim 1$.
{}From (\ref{eq:24}), (\ref{eq:25}), (\ref{eq:28}) and (\ref{eq:33}) one can
find
$\alpha=-\log{(m_w\rho)}\gg 1$. Thus, if at $r>m_w^{-1}$ the \I behaves like
(\ref{eq:32}), then at small distances $r\sim\rho$
\begin{eqnarray}\label{eq:34}
A^a_\mu(r\sim\rho)&=&\fr{2}{g} U^{ab}\bar{\eta}^b_{\mu\nu} r_\nu
\fr{\tilde{\rho}^2}{r^2(r^2+\tilde{\rho}^2)}\,\,\,\,\, , \\
\tilde{\rho}^2&=&\rho^2\left( 1+(m_w\rho)^2
\log{\left( \fr{1}{m_w\rho}\right)} \right) \,\,\,\,\, . \nonumber
\end{eqnarray}

\section{The Instanton Action}

Now we know much enough about the properties of constrained \I to find the
action for the \IC. Consider first the
contribution to the action coming from the Higgs fields. To this end
consider the second and the third terms of the action (\ref{eq:7}), but
taking the gauge field $A^a_\mu$ in the form (\ref{eq:8}). It is convenient
to divide the range of integration into two parts $|r|<L$ and $|r|>L$, where
$\rho\ll L\ll m^{-1}$. The small distances contribution up to corrections
$\sim(m\rho)^4$ reads
\begin{eqnarray}\label{eq:35}
S_{\phi,|r|<L}&=&\int_{0}^{L}\left\{  \fr{1}{2} \overline{(D^0_\mu\Phi_0)}
(D^0_\mu\Phi_0)d^4r+
 \fr{1}{2}[\overline{(D^0_\mu\delta\Phi)}
(D^0_\mu\Phi_0)+
h.c.]\right\}+ \nonumber\\ \ \nonumber\\
&+&\int_{0}^{L}\fr{m_h^2}{8\ups^2}
(\overline{\Phi_0}\Phi_0-\ups^2)^2  d^4r\,\, , \\
\ \nonumber\\
D^0_\mu&=&\partial_\mu-i\fr{g}{2}\sigma^a A^a_{\mu 0}\,\,\, .
\nonumber
\end{eqnarray}
Here $h.c.$ means hermitian conjugated.
For unperturbed fields $\Phi_0,A^a_{\mu 0}$ see eq. (\ref{eq:8}). The last
integral in (\ref{eq:35}) is easy to found explicitly. With the use of
identity $D^0_\mu D^0_\mu\Phi_0=0$  the remaining integral
in (\ref{eq:35}) can be transformed into the surface one
\begin{eqnarray}\label{eq:36}
S_{\phi,|r|<L}&=&\int_{|r|=L}d{\scriptstyle S}_\mu \left\{
\fr{1}{2}\overline{\Phi_0}D^0_\mu\Phi_0 + \fr{1}{2}
[\overline{\delta\Phi}D^0_\mu\Phi_0 +h.c.]\right\} + \nonumber\\ \ \\
&+&\fr{\pi^2\ups^2 m_h^2\rho^4}{4}
\left[ \log{\left(\fr{L}{\rho}\right) }-\fr{1}{2}\right] \,\,\, .
\nonumber
\end{eqnarray}
It is our luck, that the correction $\delta\phi$ to unperturbed Instantonic
 solution
(\ref{eq:8}) appears only in the surface integral. As we have seen
in the previous section, at $r\sim\rho$ the corrections of the
order of $(m\rho)^2$ to \I field depend on the choice of the
constraint. On the other hand, at large distances $L\gg\rho$ the
constraint independent expression (\ref{eq:25}) for $\phi$ in terms of
the McDonald functions can be used. With the expansion (\ref{eq:26})
one easily finds
\begin{eqnarray}\label{eq:37}
S_{\phi,|r|<L}&=&\pi^2\rho^2\ups^2\left[ 1-\fr{2\rho^2}{L^2} \right] -
\fr{\pi^2\ups^2 m_h^2\rho^4}{2}\left[ \log{\left( \fr{\gamma m_h
L}{2}\right)} -\fr{1}{2}\right] + \\ \ \nonumber\\
&+&\fr{\pi^2\ups^2 m_h^2\rho^4}{4} \left[ \log{\left( \fr{L}{\rho}\right)}
-\fr{1}{2}\right] \,\,\, .\nonumber
\end{eqnarray}
For the large distance contribution $(|r|>L)$ one can use the following
approximation for the field $\Phi$ covariant derivative (see eq. (\ref{eq:25}))
\bq\label{eq:38}
D_\mu^0\Phi=\left\{ \partial_\mu\left( -\fr{m_h\rho^2}{2r}K_1(m_h r)\right)
-i\sigma^a U^{ab}\bar{\eta}^b_{\mu\nu}\fr{r_\nu\rho^2}{r^4}\right\}
\left( \begin{array}{c}
0 \\ \ups
\end{array} \right)
\,\, .
\ee
The substitution of (\ref{eq:38}) and (\ref{eq:25}) into (\ref{eq:7})
leads to
\begin{eqnarray}\label{eq:39}
S_{\phi,|r|>L}&=&\int_{L}^{\infty} \left\{ \fr{\ups^2\rho^4}{8}
\left( \partial_\mu\fr{m_h}{r}K_1(m_h r) \right)^2+\fr{\ups^2\rho^4}{8}
m_h^2 \left( \fr{m_h}{r}K_1^2 \right) +\fr{3}{2}\fr{\ups^2\rho^4}{r^6}
 \right\} d^4r= \nonumber \\ \\
&=&\pi^2\rho^2\ups^2 \left\{ \fr{\rho^2}{2L^2}+\fr{m_h^2\rho^2}{4}
 \left[ \log{ \left( \fr{m_h\gamma L}{2} \right) } -1\right] \right\}
+\fr{3}{2}\fr{\pi^2\rho^4\ups^2}{L^2} \,\,\,\,\, . \nonumber
\end{eqnarray}
For integration of the squared derivative of the McDonald function it is
again useful to transform the integral to the surface one. Adding of
(\ref{eq:39}) to (\ref{eq:37}) provides us with the contribution to
the action from the Higgs fields at fixed gauge field $A^a_{\mu 0}$
(\ref{eq:8})
\bq\label{eq:40}
\delta S_{\phi}=\pi^2\rho^2\ups^2 \left\{ 1+\fr{m_h^2\rho^2}{4}
\left[ \log{\left(\fr{2}{m_h\gamma\rho}\right)}-\fr{1}{2}\right] \right\}
\,\, .
\ee
Here the first term $\pi^2\rho^2\ups^2$ is a well known result
\cite{tH,Afl} which is usually used in the calculation of total
cross-section. The new result is the second term
$\sim m^2\ups^2\rho^4\log(m\rho)$. As it is seen from eq. (\ref
{eq:40}) this term tends to increase the \I action (or, in other
words, tends to suppress the cross-section). This result is rather
natural. We have two sources for $\sim m^2\ups^2\rho^4$ correction
in the eq. (\ref{eq:40}). First is the last term of the action
(\ref{eq:7}) $\sim\lambda(\overline{\Phi}\Phi-\ups^2)^2$, which is
evidently positive. The second contribution comes from the correction
to the squared covariant derivative. But the unperturbed
solution $\Phi_0$ (see eq. (\ref{eq:8})) realizes the exact minimum
of $\overline{(D^0_\mu\Phi)}(D^0_\mu\Phi)$ term in the action
for given background
field $A^a_{\mu 0}$ and given asymptotic
$\langle\overline{\Phi}\Phi(r\rightarrow\infty)\rangle = \ups^2$. Thus the
correction to $\overline{(D^0_\mu\Phi)}(D^0_\mu\Phi)$ should also
be positive.

  Now we have to calculate the corrections to action caused by the
difference of the gauge field $A^a_\mu$ from the field of \I of pure \YM
$A^a_{\mu 0}$ (eq. (\ref{eq:8})). We will calculate such the corrections
in the large logarithm $\log (m_w\rho)$ approximation, because any of
them have no large factor $m_h^2$. For the beginning consider the second term
in the action (\ref{eq:7}) $\sim\overline{(D_\mu\Phi)}(D_\mu\Phi)$.
Two types of corrections $\sim\ups^2 m_w^2\rho^4\log (m_w\rho)$ can be found
coming from this term. The first correction is dominated by
small distances $|r|\sim\rho$. As we decided in the previous section, we call
the \I of the size $\rho$ configuration which has the asymptotic
(\ref{eq:32}). As a result of such definition we have to renormalize the
effective \I radius at small distances in accordance with (\ref{eq:34}).
Consequently, the leading term in the correction to pseudoparticle action
caused by the Higgs fields (\ref{eq:40}) should also be modified
\bq\label{eq:41}
\pi^2\rho^2\ups^2\rightarrow\pi^2\tilde{\rho}^2\ups^2=
\pi^2\rho^2\ups^2 \left( 1+(m_w\rho)^2\log\left(\fr{1}{m_w\rho}\right)
\right) \,\, .
\ee

  The second correction to the $\overline{(D_\mu\Phi)}(D_\mu\Phi)$ term
caused by the change of the gauge field $A^a_{\mu 0}$ (\ref{eq:8}) comes
from the large distances. At $r\gg\rho$ the unperturbed field $A^a_{\mu 0}$
should be replaced by the field $A^a_\mu$ (see eqs. (\ref{eq:25}) and
(\ref{eq:22})) proportional to the McDonald function. In the intermediate
range $\rho\ll r\ll m_w^{-1}$, where all the large logarithms come from, one
can use the approximation
\bq\label{eq:42}
A^a_\mu = \fr{2}{g} U^{ab}\bar{\eta}^b_{\mu\nu} r_\nu
 \left[ \fr{\rho^2}{r^4} - \fr{(m_w\rho)^2}{4r^2} \right] \,\, .
\ee
The correction induced by the second term of (\ref{eq:42})
to the $\sim \overline{(D_\mu\Phi)} (D_\mu\Phi)$ term of the action
(\ref{eq:7}) reads
\bq\label{eq:43}
\delta S_{\phi A}=-\fr{3}{2}\pi^2\ups^2 m_w^2\rho^4
\log\left(\fr{1}{m_w\rho}\right) \,\, .
\ee

  Finally, there appears a correction to the first term of the action
(\ref{eq:7}), $F^a_{\mu\nu}F^a_{\mu\nu}$. Since the unperturbed \I is an exact
extremum for the action of pure \YMC, one should consider the second variation
of the action. With the use of expression (\ref{eq:42}) it is easy to find the
corrections to the action and to $F^a_{\mu\nu}$
\begin{eqnarray}\label{eq:44}
\delta F^a_{\mu\nu}&=&\fr{(m_w\rho)^2}{g}U^{ab}
 \left[ \fr{r_\mu\bar{\eta}^b_{\nu i}r_i}{r^4}-
   \fr{r_\nu\bar{\eta}^b_{\mu i}r_i}{r^4}+
   \fr{\bar{\eta}^b_{\mu\nu}}{r^2} \right] \,\,\,\, ,
   \nonumber\\ \, \nonumber\\
\delta S_A&=&\fr{1}{4}\int_{\rho\ll r \ll m_w^{-1}}
 \delta F^a_{\mu\nu} \delta F^a_{\mu\nu} d^4r=
 3\pi^2\fr{(m_w\rho)^4}{g^2}\log\left( \fr{1}{m_w\rho}\right)
\,\, .
\end{eqnarray}

  Summing up the contributions (\ref{eq:40}),(\ref{eq:41}),(\ref{eq:43}) and
(\ref{eq:44}) we can find the action of the constrained \I (note that
$m_w=g\ups/2$)
\bq\label{eq:45}
S=\fr{8\pi^2}{g^2} \left\{ 1+\fr{(m_w\rho)^2}{2}
+\fr{m_w^2m_h^2\rho^4}{8} \left[ \log \left(
\fr{2}{m_h\gamma\rho} \right) -\fr{1}{2} \right]+
\fr{(m_w\rho)^4}{8}\log \left( \fr{const}{m_w\rho} \right)
\right\} \, ,
\ee
Where unknown $const$ is of order of $1$.

\section{Constrained \I for Heavy Higgs}\label{sec:a}

  Thus we see (\ref{eq:45}), that in the lowest order the $\rho$ dependence of
\I action is determined by the correction $\delta S\sim (m_w\rho
/g)^2$ in accordance with the results of refs. \cite{tH,Afl}. The following
correction to this result is of the relative order
$\sim (m_h\rho)^2\log (m_h\rho)$. The modern restriction on the Higgs mass does
not exclude the possibility of $m_h^2\gg m_w^2$. Therefore it is
interesting to analyze the situation when $(m_h\rho)$ is not small.

  As we have seen in Sec. \ref{sec:2}, in the spontaneously broken \YM
(\ref{eq:7}) only very small \Is may be treated semiclassically. The natural
criterion of this smallness is the inequality $m_w\rho\ll 1$. In order to get
the nontrivial solution of the equations of motion (\ref{eq:20}) we have
used the constraint for the gauge fields $A_\mu^a$. However for any given
configuration $A_\mu^a(r)$ the action (\ref{eq:7}) has a well defined exact
minimum with respect to the field $\Phi (r)$ variation. Therefore the
condition $m_h\rho\ll 1$, which we have used before seems to be excessive.

  In this Section we would consider the \I of the theory (\ref{eq:7}) for
the case of $(m_h\rho)$ not small, while $m_w\rho\ll 1$. We
concentrate our attention on the case $m_h\rho\gg 1$. There is no
principal problems to consider the case $m_h\rho\sim 1$, but for explicit
calculation of $\Phi(r)$ and \I action in this case one has to solve
numerically the nonlinear differential equation.

  Consider  the classical Higgs field configuration $\Phi(r)$ for
$(m_h\rho)\gg 1$. In the leading approximation one can neglect the
difference of the gauge field configuration $A_\mu^a$ from the unperturbed
\I $A_{\mu 0}^a$ (\ref{eq:8}). In this case the second of equations
(\ref{eq:23}) takes the form
\bq\label{eq:a1}
-\phi''-\fr{3}{r}\phi'+\fr{3}{r^2}\fr{\rho^4}{(r^2+\rho^2)^2}\phi
+\fr{m_h^2}{2} \phi(\phi^2-1)=0 \, \, .
\ee
For $m_h\rho\gg 1$ two asymptotics of solution of this equation can be
easily found. At large distances
\bq\label{eq:a2}
\phi=1-\fr{3}{(m_h r)^2}\fr{\rho^4}{(r^2+\rho^2)^2}+O(m_h^{-4}) \, \, , \,
\, m_h r\gg 1 \, \, ,
\ee
and at small distances
\begin{eqnarray}\label{eq:a3}
&\phi(0)=0 \, \, , \\
&\phi(r\ll m_h^{-1}) \sim m_h r \, \, . \nonumber
\end{eqnarray}
In the intermediate region $m_h r\sim 1$ the function $\phi(r)$ interpolates
smoothly between the solutions (\ref{eq:a2}) and (\ref{eq:a3}). In order to
find the $\phi(r)$ in the region $m_h r\sim 1$ one should solve exactly the
equation (\ref{eq:a1}). But for calculation of the leading correction to
\I action it is enough to know that $\phi$ differs very slightly from the
vacuum value $\phi =1$ at $r\sim\rho$ and at small distances $\phi(r\sim
m_h^{-1}) \sim O(1)$. Correction to the action of \I comes from the term of
the action (\ref{eq:7}) with squared covariant derivative of the Higgs field
\bq\label{eq:a4}
\delta S\approx \int \fr{1}{2}\overline{(D_\mu \Phi)}(D_\mu\Phi)\approx \int
\fr{g^2}{8} A_\mu^a A_\mu^a \overline{\Phi}\Phi \approx \fr{3\pi^2}{2}
\ups^2\rho^2 \, \, .
\ee
The corrections to this result are of the relative order $(m_h\rho)^{-1}$.

  Therefore the \I action for the case $(m_w\rho)\ll 1$, $(m_h\rho)\gg 1$ reads
\bq\label{eq:a5}
S=\fr{8\pi^2}{g^2}\left\{ 1+\fr{3}{4}(m_w\rho)^2+\ldots \right\} \,\, .
\ee
We see that for heavy Higgs even the first term of the action $S$ expansion
over the powers of $\rho^2$ differs from that for the light Higgs boson case
(\ref{eq:45}).

\section{The Baryon Number Non - Conservation}\label{sec:5}

  Before consideration of $\eps^{8/3}$ corrections to $F(\eps)\sim\log(\sigma_
{total})$ we would like to remind the reader how the first nontrivial term
$\sim\eps^{4/3}$ can be found \cite{KRT,Zakh,Por}. We will follow the paper
\cite{Diak}. The cross section for baryon number violating process accompanied
by multiple emission of massless W-bosons is given by the formula
\begin{eqnarray}\label{eq:49}
\sigma_{total}&\propto&\sum_{n}\fr{1}{n!}
\int d\rho_I d\rho_A dU_IdU_A
\mu(\rho_I)\mu(\rho_A) \cdot\\
&\cdot&\prod_{i=1}^{n}\int\fr{d^4k_i}{(2\pi)^3}\delta_+(k_i^2)
k_i^2A^{aI}_\mu(k_i)k_i^2A^{aA}_\mu(k_i)(2\pi)^4
\delta^4(P-\sum k_i) \,\, . \nonumber
\end{eqnarray}
Here $\rho_I,U_I,S_I$ and $\rho_A,U_A,S_A$ are correspondingly size,
orientation matrix and action of the \I and Antiinstanton,
$\mu(\rho_{I,A})\sim\exp(-S(\rho_{I,A}))$ are
the pseudoparticles weights (see(\ref{eq:45})),
$P_\mu=(\sqrt{s},0,0,0)$ is the total 4--momentum of produced
W-bosons. $A^{aI}_\mu$ and $A^{aA}_\mu$ are singular at $k^2=0$ parts
of the Fourier transform of the
\I and Antiinstanton field
\begin{eqnarray}\label{eq:50}
k^2 A^{aI(A)}_\mu&=&+(-) \fr{4\pi^2i\rho^2U^{ab}\bar{\eta}^b_{\mu\nu}k_\nu}
{g} \ .
\end{eqnarray}
Only the residue of the \I (Antiinstanton) field Fourier transform at $k^2=0$
is needed for the calculation of cross-section (\ref{eq:49}). Now it is clear,
why do we have defined the constrained \I by the formula (\ref{eq:32}). The
field (\ref{eq:32}) corresponds to a massive particle, therefore its Fourier
transform has a pole at $k^2+m_w^2=0$ ($k$ is the Euclidean momenta), not at
$k^2=0$. But the residue at this pole is exactly the same as for the
unperturbed \I (\ref{eq:50}).

  One can introduce a distance between the \I and Antiinstanton $R_\mu$
in the usual way \cite{KRT,Zakh,Por}
\bq\label{eq:51}
(2\pi)^4\delta^4(P-\sum k_i)=\int d^4R\exp\left\{i(PR)-i\sum(k_iR)
\right\} \,\, ,
\ee
\bq\label{eq:52}
\int\fr{d^4k}{(2\pi)^3}\delta_+(k^2)e^{-i(kR)}=\fr{1}
{4\pi^2[\vec{R}^2-(R_0-i0)^2]} \,\, .
\ee
After that the factorized integrals over 4-momentum of produced
W-bosons sum up to the exponent:
\begin{eqnarray}\label{eq:53}
\sigma_{total}&=&\int d^4Re^{i(PR)}
d\rho_I d\rho_A dU_IdU_A
\mu(\rho_I)\mu(\rho_A)\cdot \nonumber\\
&\cdot&\exp\left\{ -\fr{8\pi^2}{g^2}
\fr{4D\rho_1^2\rho_2^2}{[\vec{R}^2-(R_0-i0)^2]^2} \right\} \,\,\, , \\
D&=&\bar{\eta}^a_{\mu\alpha}(U_I^T U_A)^{ab} \eta^b_{\mu\beta}
\fr{R_\alpha R_\beta}{R^2} \, \in \, (-3,1) \  .\nonumber
\end{eqnarray}
The integration over $R_\mu$ and $U_{I,A}$ is performed by the steepest
descent method with the saddle-point values given by
\bq\label{eq:54}
D=-3\, , \, R_0=-i\left( \fr{384\pi^2\rho_I^2\rho_A^2}
{g^2\sqrt{s}} \right)^{1/5} \, , \, \vec{R} =0 \,\, .
\ee
The integrals over $\rho_I,\rho_A$ are also performed by the steepest
descent method. In the leading order one should neglect the $\rho^4$
term in the \I action(\ref{eq:45}). The following saddle-point values
are easy to find
\bq\label{eq:55}
(\rho m_w)^2=\fr{3}{2} \eps^{4/3} \,\, ,\,\, (Rm_w)^2=6\eps^{2/3}
\,\, ,
\ee
where $\eps=\sqrt{s}/E_0 , E_0=\sqrt{6}\pi m_w/\alpha =
4\sqrt{6}\pi^2 m_w/g^2$. After all, in the leading semiclassical
approximation the total cross-section is given by \cite{KRT,Zakh,Por}
\bq\label{eq:56}
\sigma_{total}\propto\exp\left[ \fr{4\pi}{\alpha}F(\eps)\right]
\approx\exp\left[\fr{4\pi}{\alpha}\left( -1+\fr{9}{8}
\eps^{4/3}\right)\right] \,\,\, .
\ee
As can be seen from (\ref{eq:55}) the function $F(\eps)$ is naturally
expanded in powers of $\eps^{2/3}$. The following $\eps^2$ term
\cite{Khos,DiPe,Muel,ArMa} reads
\bq\label{eq:57}
F_2=-\fr{9}{16}\eps^2 \,\,\,\,\, .
\ee
There are three effects which contribute to this value. They are:
\begin{enumerate}
\item the quantum correction to the classical cross-section
(\ref{eq:56}),
\item the correction which comes from the slight modification of the
classical result due to the finite W-boson mass,
\item the correction due to the multiple production of massless Higgs bosons.
\end{enumerate}

  Going further the authors of \cite{Diak} have calculated the second
order quantum correction to $F$ in the limit of massless W-bosons
\bq\label{eq:58}
F_{8/3}^{qu}=+\fr{3}{16} \eps^{8/3} \log \left( \fr{c}{\eps} \right)
\,\,\,\, .
\ee

  In the previous section (see eq. (\ref{eq:45})) we have calculated
the correction $\sim (m\rho)^4\log (m\rho)$ to the single-Instanton action.
With the saddle point values (\ref{eq:55}) it is seen immediately,
that the modification of Instantonic measure due to this term
provides the correction to $F(\eps)$ of the same order as
(\ref{eq:58})
\bq\label{eq:59}
F_{8/3}^{\mu} =-\fr{3}{16}\eps^{8/3} \left\{
\left(\fr{m_h}{m_w}^2 \right) \left[ \log\left( \fr{1}{\eps}\right)
+ \fr{3}{2}\log\left( \fr{2\sqrt{2}m_w}{\gamma\sqrt{3}m_h}\right)
-\fr{3}{4}\right] + \log\left(\fr{1}{\eps}\right) \right\} \,\, .
\ee

  A number of effects exist which may lead to a correction of the
order of $\eps^{8/3}$ to the function $F(\eps)$, such as further
corrections due to a finite W mass, or quantum corrections to multiple Higgs
emission. But none of them have either $\log(\eps)$, or $(m_h/m_w)^2$
enhancement. We have found only one effect which leads to a correction
comparable with (\ref{eq:59}) and (\ref{eq:58}). This is the correction
caused by the finite Higgs mass $m_h$ to the classical multiple Higgs
production. The account of the multiple Higgs boson production can
be done in the similar way, just like the total cross-section for baryon
number violating process accompanied by
the classical multiple W-s production (\ref{eq:56}) was found. One should
only replace in the formula (\ref{eq:49}) the Fourier transform of the gauge
fields $A^{aI}_\mu(k), A^{aA}_\mu(k)$ by that of Higgs field (\ref{eq:25})
\bq\label{eq:60}
\Phi (k)=\fr{2\pi^2\rho^2}{k^2-m_h^2}\left( \begin{array}{c}
0 \\ \ups
\end{array} \right) \ .
\ee
Here we work in the Minkowski space-time. As far as we want to take into
account the finite mass effect, the multipliers $k_i^2$ in (\ref{eq:49}) at the
field Fourier transform should be also replaced by $(k_i^2-m_h^2)$. Finally,
one has to modify the expression (\ref{eq:52})
\begin{eqnarray}\label{eq:61}
\int\fr{d^4 k}{(2\pi)^3}& & \delta_+ (k^2-m^2)e^{-ikR}=
\ \\
&=&{\displaystyle \fr{1}{4\pi^2}\fr{m}{\sqrt{\vec{R}^2-R_0^2}}}
\ K_1(m\sqrt{\vec{R}^2-R_0^2}) \ , \nonumber \\
& &R_0^2<\vec{R}^2 \ . \nonumber
\end{eqnarray}
After taking into account the multiple production of massive Higgs bosons
in the leading classical approximation the cross-section reads
\bq\label{eq:62}
\sigma_{total}=\sigma_w \, \exp \left\{ \pi^2\ups^2\rho_1^2\rho_2^2
\fr{m_h}{\sqrt{\vec{R}^2-R_0^2}} K_1(m_h\sqrt{\vec{R}^2-R_0^2})
\right\} \ ,
\ee
where $\sigma_w$ is the cross-section for multiple W-s production. If one
leaves in eq. (\ref{eq:62}) the first term of the McDonald function asymptotic
at $m_h\sqrt{\vec{R}^2-R_0^2} \ll 1$ only, then all the Higgs mass $m_h$
dependence
disappears and one gets the same $\sim\eps^2$ correction to $F(\eps)$, as
was used in \cite{Khos,DiPe,Muel,ArMa}. The following term of the McDonald
function $K_1$ in (\ref{eq:62}) expansion (see eqs. (\ref{eq:26}) and
(\ref{eq:55})) gives
\bq\label{eq:63}
F_{8/3}^{m_h}=-\fr{3}{32} \eps^{8/3} \left( \fr{m_h}{m_w} \right)^2
\left[ \log\left( \fr{1}{\eps} \right) + 3\log \left( \sqrt{\fr{2}{3}}
\fr{m_w}{\gamma m_h} \right) +\fr{3}{2} \right] \ .
\ee
Summing up of (\ref{eq:56}), (\ref{eq:57}), (\ref{eq:58}), (\ref{eq:59}) and
(\ref{eq:63}) provides us with the result
\bq\label{eq:64}
F=-1+\fr{9}{8}\eps^{4/3}-\fr{9}{16}\eps^2-\fr{9}{32}
\left(\fr{m_h}{m_w}\right)^2 \eps^{8/3}
\log\left[ \fr{1}{3\eps}\left(\fr{2m_w}{\gamma m_h}\right)^2\right]
+O(\eps^{8/3}) \ .
\ee
Here the unknown correction behaves like $\eps^{8/3}$, but have
neither $\log(\eps)$, nor $(m_h/m_w)^2$ enhancement.

  It is interesting to discuss the range of applicability of the expression
(\ref{eq:64}). As we have said in the Introduction, one can use the
approximate expression (\ref{eq:64}) instead of the exact $F(\eps)$ if at
least the contribution of the last term $\sim\eps^{8/3}$ is smaller than the
$\eps^2$ contribution. The result of \cite{Diak} still may be interpreted as
the indication of applicability of the $F(\eps)$ expansion in powers of $\eps$
up to $\eps\sim 1$. Unfortunately, as far as the final result (\ref{eq:64}) is
considered $\eps^{3/8}$ and $\eps^2$ terms became equal at $\eps=0.3\div0.12$
for $m_h=2\div 3m_w$. Only for this region $\eps<0.3\div0.12$ one can believe
in the applicability of the expansion (\ref{eq:64}).

  Moreover, the more severe restriction for applicability of the expression
(\ref{eq:64}) can be formulated. As we have seen, at sufficiently small $\eps$
the correction $\sim\eps^{8/3}$ tends to decrease the total cross-section. As
we would argue, this is not an occasion. For example, the correction
$F^{m_h}_{8/3}$ (\ref{eq:63}) arises when we take into account the finite
Higgs boson mass. This correction should suppress the cross-section, because a
phase space for massive particle is smaller then the phase space available for
massless one. The correction of the order of $m_h^2m_w^2\rho^4$ to the \I
action (\ref{eq:45}) should also decrease the cross-section as we have argued
in the previous section (see the discussion after eq.(\ref{eq:40})). Thus, one
can neglect the following terms in the expansion (\ref{eq:64}) for $F(\eps)$
($\sim\eps^{10/3}, \ \sim\eps^4$ et.c.) only if the last term $\sim\eps^{8/3}
(m_h/m_w)^2\log(\eps)$ is negative. This means
\begin{eqnarray}\label{eq:65}
\eps < \fr{1}{3} \left( \fr{2}{\gamma} \fr{m_w}{m_h} \right)^2  \ .
\end{eqnarray}

  Since of at $m_h\gg m_w$ the role of corrections dependent on the Higgs mass
in (\ref{eq:64}) increase drastically, it seems very attractive
to try to take into account all the $(m_h/m_w)^2$ corrections exactly. In
the previous Section we have found the correction to \I action for the case
$m_w\rho\ll 1\ll m_h\rho$. Repeating all the considerations, which have lead
us to the result (\ref{eq:56}), with the new \I action (\ref{eq:a5}) we get
\begin{eqnarray}\label{b1}
&F(\eps)= -1+{\displaystyle \fr{9}{8}\left( \fr{2}{3}\right)^{2/3}\eps^{4/3}}
+\ldots \,\, , \\
&\eps\ll 1\,\, , \,\, \eps{\displaystyle \left(\fr{m_h}{m_w} \right)^{3/2}}
 \gg 1 \,\, . \nonumber
\end{eqnarray}
For the range of applicability of this result we have used the same
inequality $m_w\rho\ll 1\ll m_h\rho$ expressed in terms of the saddle point
values (\ref{eq:55}). Unfortunately it seems a severe problem even to
estimate the corrections to this result coming from multiple heavy Higgs
production. We can describe satisfactorily only the Euclidean classical
configurations. But in order to describe the Higgs boson production one
should reach the far Minkowski point $m_h^2+k^2=0$.

\section{Massive $\phi^4$ theory}\label{sec:Phi}

Consider the theory with Euclidean action
\bq\label{eq:66}
S=\fr{1}{g}\int\left[ \fr{1}{2}(\partial_\mu \phi)^2+\fr{1}{2}m^2\phi^2
-\fr{1}{4!}\phi^4\right] d^4r \ ,
\ee
where $\phi$ is a real scalar field. In the absence of mass $(m=0)$ where
exist the \I of an
arbitrary size $\rho$. The explicit formula for \I and its dilatational zero
mode reads
\bq\label{eq:67}
\phi_0=\fr{4\sqrt{3}\rho}{\rho^2+r^2} \ \ , \ \
\fr{\partial\phi_0}{\partial\rho}\sim\fr{r^2-\rho^2}{(r^2+\rho^2)^2} \ .
\ee
Simple scale transformation $\phi(x) \rightarrow a\phi(ax)$ ensures one that
the exact extremum of the action of the massive theory is the zero size
\I only. We will discuss a small size \Is $m\rho\ll 1$. Similarly to
the case of spontaneously broken \YM one has to introduce a constraint to the
functional integral in order to integrate over the \I radius $\rho$
\bq\label{eq:68}
\int\delta(\langle \phi-\phi_I |w\rangle) \
d\langle \phi-\phi_I |w\rangle \ .
\ee
After that the equation of motion for Instantonic configuration takes the form
\bq\label{eq:69}
-\triangle\phi -\fr{1}{6}\phi^3 +m^2\phi=\zeta w \ ,
\ee
where $\zeta$ is the Lagrange multiplier. As well as for the \YMC, there is no
best way to choose the constraint $w$, and therefore there is no the  unique
choice of the Instantonic configuration. One can state only that the
constraint $w(r)$ should not be orthogonal to the dilatational zero mode
$\partial\phi_0/\partial\rho$ and should not be very close to
$\partial\phi_0/\partial\rho$. We will assume, that $w(r)$ is a spherically
symmetric function, which goes to zero very fast at $r>\rho$. In this case the
solution of eq. (\ref{eq:69}) at $r\ll m^{-1}$ and $r\gg\rho$ takes the form
\begin{eqnarray}\label{eq:70}
& &\phi\approx\phi_0 +\delta\phi \ \ , \ \ r\ll m^{-1} \ \ , \\
& &\phi\approx 4\sqrt{3} \rho\fr{m}{r} K_1(mr) \ \ , r\gg\rho \ . \nonumber
\end{eqnarray}
Here $\delta\phi\sim m^2$ is a correction, which can be easily found in the
same way as it was done for the \YM (see (\ref{eq:24}), (\ref{eq:28}),
(\ref{eq:30})). For us it is only important that in the intermediate region
$\rho\ll r\ll m^{-1}$ (see (\ref{eq:26}))
\bq\label{eq:71}
\delta\phi= 2\sqrt{3} \rho m^2 \left[ \log\left( m\gamma\fr{r}{2}
\right) -\fr{1}{2} \right] \ .
\ee

  The finite mass correction to the action is of the form $\triangle
S\sim(m\rho)^2 \log(m\rho)/g$ \cite{Afl,FY}. The only term in the action
(\ref{eq:66}), which may provide us with the large logarithm is
\bq\label{eq:72}
\triangle S\approx \fr{1}{g}\int\fr{m^2\phi^2}{2}d^4r
\approx\fr{1}{g}\int_{0}^{m^{-1}}\fr{m^2\phi_0^2}{2}d^4r
\approx\fr{48\pi^2(m\rho)^2}{g}\log\left(\fr{1}{m\rho}\right) \ .
\ee
To our surprise this result occurs to be twice larger than usually used
(see e.g. \cite{ShM}) result of the paper \cite{Afl}.

  For more accurate calculation of the correction to the \I action it is useful
to divide the range of integration into two parts $r<R$ and $r>R$
$(\rho\ll R\ll m^{-1})$ as we have done for the \YMC. The short range
contribution reads
\begin{eqnarray}\label{eq:73}
S_{r<R}&=&\fr{1}{g}\left\{ \int_{0}^{R}\left[ \fr{(\partial_\mu\phi_0)^2}{2}
+\fr{m^2}{2}\phi_0^2 - \fr{\phi_0^4}{4!}\right] +\int_{|r|=R} \delta\phi
\partial_\mu\phi_0 ds_\mu \right\}= \nonumber\\
 \ \\
&=&\fr{16\pi^2}{g}\left\{ 1-6\left( \fr{\rho}{R} \right)^2 +3(m\rho)^2
\left[ 2\log\left( \fr{2}{m\gamma R}\right) -\log\left(\fr{\rho}{R}\right)
+\fr{1}{2} \right] \right\} \ . \nonumber
\end{eqnarray}
The large distance contribution reads
\begin{eqnarray}\label{eq:74}
S_{r>R}&=&\fr{48\rho^2}{g}\int_{R}^{\infty} \left[ \fr{1}{2}
\left( \partial_\mu\fr{m}{r} K_1(mr)\right)^2 +\fr{m^2}{2}
\left( \fr{m}{r} K_1(mr)\right)^2 \right] = \nonumber\\
\ \nonumber\\
&=&-\fr{24\rho^2}{g}\int_{|r|=R} \fr{mK_1(mr)}{r}\partial_\mu
\fr{mK_1(mr)}{r}ds_\mu = \\
\ \nonumber\\
&=&96\pi^2\left(\fr{\rho}{R}\right)^2 +48\pi^2(m\rho)^2\left[
\log\left(\fr{m\gamma R}{2}\right)-1 \right] \ . \nonumber
\end{eqnarray}
Summing up (\ref{eq:73}) and (\ref{eq:74}) we get the action of
constrained \I in the massive $\phi^4$ theory
\bq\label{eq:75}
S=\fr{16\pi^2}{g}\left\{ 1+3m^2\rho^2 \left[ \log\left(
\fr{2}{\gamma m\rho}\right) -\fr{1}{2}\right] \right\} \ .
\ee
Corrections to this formula are of the order of $(m\rho)^4/g$ and generally
speaking depends on the constraint.

{\large\bf Asknowledgements.} Author is grateful to M.E.\,\,Pospelov for taking
part at the early \,stage\, of \,the\, work\, and\, to\, V.L.\,\,Chernyak,\,
 M.V.\,\,Mostovoy,\, O.P.\,\,Sushkov \,\,and A.S.\,\,Yelkhovsky for numerous
helpful discussions. The kind interest and discussions of D.I.\,\,Diakonov,
V.Yu.\,\,Petrov and M.V.\,\,Polyakov are also greatly appreciated.

\end{document}